\documentclass[a4paper,onecolumn,11pt,accepted=2024-10-22]{quantumarticle}
\pdfoutput=1
\usepackage[utf8]{inputenc}
\usepackage[english]{babel}
\usepackage[T1]{fontenc}
\usepackage{hyperref}
\usepackage{graphicx}
\usepackage[a4paper]{geometry}
\usepackage{amsmath,amssymb}
\usepackage{braket}
\usepackage{color}
\usepackage{enumerate}
\usepackage{empheq}

\usepackage{tikz}
\usepackage{lipsum}

\begin{document}

\title{Comment on ``Multivariable quantum signal processing (M-QSP): prophecies of the two-headed oracle''}

\author{Hitomi Mori}
\affiliation{Graduate School of Engineering Science, Osaka University, 1-3 Machikaneyama, Toyonaka, Osaka 560-8531, Japan}
\email{hmori.academic@gmail.com}
\author{Kaoru Mizuta}
\affiliation{Department of Applied Physics, The University of Tokyo, Hongo 7-3-1, Bunkyo, Tokyo 113-8656, Japan}
\affiliation{Center for Quantum Computing, RIKEN, Hirosawa 2-1, Wako, Saitama 351-0198, Japan}
\author{Keisuke Fujii}
\affiliation{Graduate School of Engineering Science, Osaka University, 1-3 Machikaneyama, Toyonaka, Osaka 560-8531, Japan}
\affiliation{Center for Quantum Information and Quantum Biology, Osaka University, 1-3 Machikaneyama, Toyonaka, Osaka 560-8531, Japan}
\affiliation{Center for Quantum Computing, RIKEN, Hirosawa 2-1, Wako, Saitama 351-0198, Japan}
\affiliation{Fujitsu Quantum Computing Joint Research Division at QIQB,
Osaka University, 1-2 Machikaneyama, Toyonaka 560-0043, Japan}
\maketitle

\begin{abstract}
  Multivariable Quantum Signal Processing (M-QSP) \cite{Rossi2022multivariable} is expected to provide an efficient means to handle polynomial transformations of multiple variables simultaneously. However, we identified several inconsistencies in the main theorem, where necessary and sufficient conditions for achievable polynomials are provided, and its proof in Ref. \cite{Rossi2022multivariable}. Moreover, a counterexample to the conjecture in Ref. \cite{Rossi2022multivariable}, based on which the main theorem is constructed, is presented in Ref. \cite{németh2023variants}, meaning the requirement of the conjecture should be included as a condition in the theorem. Here we note our observations and propose the revised necessary conditions for M-QSP. We also show that these necessary conditions cannot be sufficient conditions, and thus some additional condition on top of these revisions is essentially required for the complete M-QSP theorem. 
\end{abstract}

\section{Introduction}
Quantum algorithms have been successfully proven an advantage over classical algorithms in many computational tasks, including factoring \cite{shor_1994}, Hamiltonian simulation \cite{lloyd_1996}, and matrix inversion \cite{harrow_2009}. More recently, techniques called quantum signal processing (QSP) \cite{low_2016} and quantum singular value transform (QSVT) \cite{Gily_n_2019} were proposed and successfully unified these quantum algorithms within the framework. Specifically, QSP performs a polynomial transformation of an element of a signal operator represented as a single qubit gate, while QSVT is an extension of QSP to a larger dimension via qubitization, and notably these techniques provide optimal complexity in many cases.

Multivariable quantum signal processing (M-QSP) \cite{Rossi2022multivariable} is proposed as a multivariable generalization of QSP. It performs multivariable polynomial transform using multiple kinds of signal operators represented as $X$-rotation, which correspond to each variable. The original paper of M-QSP provided necessary and sufficient conditions on polynomials implementable with M-QSP in its main theorem (Theorem 2.3). However, we identified several inconsistencies in the theorem and its proof. Separately, a counterexample to the key conjecture (Conjecture 2.1), which is about the coefficients of multivariable polynomials and on which the theorem is constructed, is found in Ref. \cite{németh2023variants}, meaning the requirement of the conjecture should be included as a condition in the theorem. In this paper, we note our observations and propose the revised necessary conditions for M-QSP. We also show that these necessary conditions cannot be sufficient conditions, and thus some additional condition on top of these revisions is essentially required for the complete M-QSP theorem.

The rest of this paper is organized as follows. In Section \ref{sec1} and Section \ref{sec2}, we explain the inconsistencies in Theorem 2.3 of Ref. \cite{Rossi2022multivariable} and its proof, respectively. We propose the revised theorem in Section \ref{sec3}, followed by the conclusion in Section \ref{sec4}.

\section{Inconsistency in Theorem 2.3}\label{sec1}
In this section we show inconsistencies observed in Theorem 2.3 in Ref. \cite{Rossi2022multivariable}.\\

\noindent{\bf Theorem 2.3 in Ref.~\cite{Rossi2022multivariable}:} \textit{Let $n \in \mathbb{N}$. There exist $\Phi=\left\{\phi_0, \phi_1, \cdots, \phi_n\right\} \in \mathbb{R}^{n+1}$ and $s=\{s_1,\cdots,s_n\}\in\{0,1\}^n$ such that for all $(a, b) \in \mathbb{T}^2$:
\begin{align}
    U_{(s, \Phi)}(a, b)=e^{i \phi_0 \sigma_z} \prod_{k=1}^n A^{s_k}\left(a\right) B^{1-s_k}\left(b\right) e^{i \phi_k \sigma_z}=\begin{pmatrix}
        P & Q \\
-Q^* & P^*
    \end{pmatrix},
\end{align}
where $A=\mathbb{I}\left(a+a^{-1}\right) / 2+\sigma_x\left(a-a^{-1}\right) / 2$ and $B=\mathbb{I}\left(b+b^{-1}\right) / 2+\sigma_x\left(b-b^{-1}\right) / 2$ if and only if $P, Q \in \mathbb{C}[a, b]$ (Laurent polynomials) and}
\begin{enumerate}[(i)]
\item \textit{$\operatorname{deg}(P) \preccurlyeq(m, n-m)$ and $\operatorname{deg}(Q) \preccurlyeq(m, n-m)$ for $m=|s|$ the Hamming weight of $s$.}
\item \textit{$P$ has parity- $n(\bmod 2)$ under $(a, b) \mapsto\left(a^{-1}, b^{-1}\right)$ and $Q$ has parity- $(n-1)(\bmod 2)$ under $(a, b) \mapsto\left(a^{-1}, b^{-1}\right)$.}
\item \textit{$P$ has parity $m(\bmod 2)$ under $a \mapsto-a$ and parity $m-n(\bmod 2)$ under $b \mapsto-b$ and $\mathrm{Q}$ has parity $m-1(\bmod 2)$ under $a \mapsto-a$ and parity $n-m-1(\bmod 2)$ under $b \mapsto-b$.}
\item \textit{For all $(a, b) \in \mathbb{T}^2$ the relation $|P|^2+|Q|^2=1$ holds.}
\item \textit{The statement of Conjecture 2.1 holds as given.}\\\par
\end{enumerate}
\noindent
{\bf Conjecture 2.1 in Ref.~\cite{Rossi2022multivariable}:} \textit{Given a unitary matrix satisfying the conditions (i)-(iv) of Theorem 2.3, the single-variable Laurent polynomial coefficients of $P,Q$ satisfy one or both of the following relations for $\phi_A,\phi_B\in\mathbb{R}$:
\begin{align}
    P_{d_A}(b)=e^{i\phi_A}Q_{d_A}(b),
    P_{d_B}(a)=e^{i\phi_B}Q_{d_B}(a).
\end{align}
Here $d_A$ and $d_B$ are the maximal positive degree of $a,b$ respectively appearing in $P,Q$, and $P_{d_A}(b)$ denotes the single variable Laurent polynomial (in $b$) coefficient of the $a^{d_A}$ term of $P$, and analogously for the other terms.}\\\par

Here we show the inconsistency among the statements (i)-(iv). According to Theorem 2.3 (i), $P$ and $Q$ can be expressed as
\begin{align}
    &P(a,b)=\sum_{j=-m}^m\sum_{k=-(n-m)}^{n-m}P_{jk}a^jb^k,\\
    &Q(a,b)=\sum_{j=-m}^m\sum_{k=-(n-m)}^{n-m}Q_{jk}a^jb^k.
\end{align}
\\
However, from Theorem 2.3 (iii), the possible highest and lowest order term of $Q$ are restricted as
\begin{align}
    Q(a,b)=\sum_{j=-(m-1)}^{m-1}\sum_{k=-(n-m-1)}^{n-m-1}Q_{jk}a^jb^k,    
\end{align}
suggesting $\operatorname{deg}(Q) \preccurlyeq(m-1, n-m-1)$.
\par
Theorem 2.3 (iv) can be rewritten as
\begin{align}
    \sum_{j,k}|P_{jk}|^2+\sum_{(j,k)\ne(j',k')}P_{jk}P^*_{j'k'}e^{i(j-j')\theta_a}e^{i(k-k')\theta_b}
    +\sum_{j,k}|Q_{jk}|^2+\sum_{(j,k)\ne(j',k')}Q_{jk}Q^*_{j'k'}e^{i(j-j')\theta_a}e^{i(k-k')\theta_b}=1,
\end{align}
where $a=\exp(i\theta_a)$ and $b=\exp(i\theta_b)$ are substituted. Because products of orthogonal functions form orthogonal functions, $\exp(il_a\theta_a)\exp(il_b\theta_b)$ terms, where $(l_a,l_b)\equiv(j-j',k-k')\ne(0,0)$, need to vanish. Thus, the condition of Theorem 2.3 (iv) is equivalent to
\begin{align}
    \sum_{j,k}\left(|P_{jk}|^2+|Q_{jk}|^2\right)=1
\end{align}
and
\begin{align}    
    \sum_{j,k}\left(P_{jk}P^*_{j-l_a~k-l_b}+Q_{jk}Q^*_{j-l_a~k-l_b}\right)=0.\label{cond2}
\end{align}
In the case of $(l_a,l_b)=(2m,2(n-m))$, the only possible choice is $j=m$ and $k=n-m$, and hence  Eq.~\eqref{cond2} leads to
\begin{align}
    P_{m~n-m}P^*_{-m~-(n-m)}=0,\label{cond2_highest}
\end{align}
as $Q$ does not have the $(m,n-m)$ order term. According to Theorem 2.3 (ii), we have
\begin{align}
    P(a^{-1},b^{-1})=(-1)^nP(a,b),
\end{align}
which can be written down as
\begin{align}
    \sum_{j,k}P_{jk}a^{-j}b^{-k}&=(-1)^n\sum_{j,k}P_{jk}a^jb^k\\
    &=(-1)^n\sum_{j,k}P_{-j-k}a^{-j}b^{-k}.
\end{align}
Comparing the highest term on both sides, we get the condition on the coefficient:
\begin{align}
    P_{-m~-(n-m)}=(-1)^nP_{m~n-m}.
\end{align}
Plugging this into Eq. \eqref{cond2_highest}, we get
\begin{align}
    (-1)^n|P_{m~n-m}|^2=0
\end{align}
and this implies the $(m,n-m)$ order term of $P$ vanishes. Repeating the same thing for other combinations of  $(l_a,l_b)$, it turns out all coefficients other than the constant term vanish after all
\footnote{This can be seen by first considering $(l_a,l_b)=(2m,2(n-m)-2),(2m,2(n-m)-4),...,(2m,-2(n-m))$, where we find all $P_{m*}$ become 0. The same goes for $P_{*n-m}$. Then we next consider $(l_a,l_b)=(2m-2,2(n-m)-2),(2m-2,2(n-m)-4),...,(2m-2,-2(n-m))$ and we find $Q_{m-1*}$ vanish. Continuing this we find that all the coefficients other than the constant term vanish.}
, which contradicts with Theorem 2.3 (i).

\section{Inconsistency in the proof of Theorem 2.3}\label{sec2}
In this section we show inconsistencies observed in the proof of Theorem 2.3 in Ref. \cite{Rossi2022multivariable}.
\subsection{Regarding the proof of ``$\Rightarrow$'':}
When we assume the $(n-1)$th order (the $(m-1)$th order in $a$) sequence $U_{(\tilde{s},\tilde{\Phi})}$ satisfies the conditions, we get
\begin{align}
    &U_{(\tilde{s},\tilde{\Phi})}=\begin{pmatrix}\tilde{P}&\tilde{Q} ,\\
    -\tilde{Q}^*&\tilde{P}^*\end{pmatrix} ,\\
    &\tilde{P}(a^{-1},b^{-1})=(-1)^{n-1}\tilde{P}(a,b), \\
    &\tilde{Q}(a^{-1},b^{-1})=(-1)^{n-2}\tilde{Q}(a,b), \\
    &\tilde{P}(-a,b)=(-1)^{m-1}\tilde{P}(a,b), \\
    &\tilde{Q}(-a,b)=(-1)^{m-2}\tilde{Q}(a,b).
\end{align}
Then for the $n$th order (the $m$th order in $a$), which can be obtained by applying $Ae^{i\phi_n\sigma_z}$ from the right side, we get
\begin{align}
    {U}_{(s,\Phi)}=
    \begin{pmatrix}
    e^{i \phi_n}\left(\frac{a+a^{-1}}{2}\tilde{P}+\frac{a-a^{-1}}{2} \tilde{Q}\right)&e^{-i \phi_n}\left(\frac{a-a^{-1}}{2} \tilde{P}+\frac{a+a^{-1}}{2} \tilde{Q}\right)\\
    \cdot&\cdot
    \end{pmatrix}.
\end{align}
Therefore, if we consider each element in ${U}_{(s,\Phi)}$, we have
\begin{align}
    {P}(a^{-1},b^{-1})&=e^{i\phi_n}\left(\frac{a+a^{-1}}{2}\tilde{P}(a^{-1},b^{-1})-\frac{a-a^{-1}}{2} \tilde{Q}(a^{-1},b^{-1})\right)\nonumber\\
    &=(-1)^{n-1}P(a,b)\label{ind1} ,
\end{align}    
\begin{align}
    {Q}(a^{-1},b^{-1})&=e^{-i\phi_n}\left(-\frac{a-a^{-1}}{2}\tilde{P}(a^{-1},b^{-1})+\frac{a+a^{-1}}{2} \tilde{Q}(a^{-1},b^{-1})\right)\nonumber\\
    &=(-1)^{n-2}Q(a,b)\label{ind2},
\end{align}   
\begin{align}
    {P}(-a,b)&=e^{i\phi_n}\left(-\frac{a+a^{-1}}{2}\tilde{P}(-a,b)-\frac{a-a^{-1}}{2} \tilde{Q}(-a,b)\right)\nonumber\\
    &=(-1)^{m}e^{i\phi_n}\left(\frac{a+a^{-1}}{2}\tilde{P}(a,b)-\frac{a-a^{-1}}{2} \tilde{Q}(a,b)\right)\label{ind3},
\end{align}  
\begin{align}
    {Q}(-a,b)&=e^{-i\phi_n}\left(-\frac{a-a^{-1}}{2}\tilde{P}(-a,b)-\frac{a+a^{-1}}{2} \tilde{Q}(-a,b)\right)\nonumber\\
    &=(-1)^{m-1}e^{-i\phi_n}\left(\frac{a-a^{-1}}{2}\tilde{P}(a,b)-\frac{a+a^{-1}}{2} \tilde{Q}(a,b)\right).\label{ind4}
\end{align} 
However, to prove Theorem 2.3 (ii), Eq. \eqref{ind1} and Eq. \eqref{ind2} need to be $(-1)^{n}P(a,b)$ and $(-1)^{n-1}Q(a,b)$, and to prove Theorem 2.3 (iii), Eq. \eqref{ind3} and Eq. \eqref{ind4} be $(-1)^{m}P(a,b)$ and $(-1)^{m-1}Q(a,b)$. Therefore, the statement is not proven, as it does not hold for the $n$th order (the $m$th order in $a$).

\subsection{Regarding the proof of ``$\Leftarrow$'':}
After we assume the statement for the $(n-1)$th order (the $(m-1)$th order in $a$) sequence $U_{(\tilde{s},\tilde{\Phi})}$ is true (i.e. polynomial satisfies the conditions (i)-(v) $\Rightarrow$ $U_{(\tilde{s},\tilde{\Phi})}$ exists), we assume the $n$th order (the $m$th order in $a$) $P,Q$ satisfy the conditions. Thus,
\begin{align}
    &\operatorname{deg}(P) \preccurlyeq(m, n-m),\\
    &\operatorname{deg}(Q)\preccurlyeq(m, n-m),\\
    &P(a^{-1},b^{-1})=(-1)^nP(a,b),\\
    &Q(a^{-1},b^{-1})=(-1)^{n-1}Q(a,b),\\
    &P(-a,b)=(-1)^{m}P(a,b),\\
    &Q(-a,b)=(-1)^{m-1}Q(a,b).
\end{align}
Assuming $P_{d_A}(b)=e^{2i\phi_n}Q_{d_A}(b)$ from Conjecture 2.1 holds, where $d_A$ is the highest degree of $a$, we define
\begin{align}
    \begin{pmatrix}
        \tilde{P}&\tilde{Q}\\
        -\tilde{Q}^*&\tilde{P}^*
    \end{pmatrix}
    &\equiv
    \begin{pmatrix}
        P&Q\\
        -Q^*&P^*
    \end{pmatrix}e^{-i\phi_n\sigma_z}A^\dag\nonumber\\
    &=
    \begin{pmatrix}
        \frac{a+a^{-1}}{2}e^{-i\phi_n}P-\frac{a-a^{-1}}{2}e^{i\phi_n}Q&-\frac{a-a^{-1}}{2}e^{-i\phi_n}P+\frac{a+a^{-1}}{2}e^{i\phi_n}Q\\
        \cdot&\cdot\label{tilde}
    \end{pmatrix}.
\end{align}
Because the highest terms cancel out in Eq. \eqref{tilde},
\begin{align}
    &\operatorname{deg}(\tilde{P}) \preccurlyeq(m-1, n-m),\\
    &\operatorname{deg}(\tilde{Q})\preccurlyeq(m-1, n-m).  
\end{align}
Then we want to apply the assumption for the $(n-1)$th order (the $(m-1)$th order in $a$) to $\tilde{P},\tilde{Q}$ to show there exist the solution $\Phi=(\tilde{\Phi},\phi_n)$ and $s=(\tilde{s},1)$ but cannot because they do not satisfy the conditions (ii)-(iii).
\begin{align}
    &\tilde{P}(a^{-1},b^{-1})=(-1)^n\tilde{P}(a,b)\\
    &\tilde{Q}(a^{-1},b^{-1})=(-1)^{n-1}\tilde{Q}(a,b)\\
    &\tilde{P}(-a,b)=(-1)^{m-1}\frac{a+a^{-1}}{2}e^{-i\phi_n}P(a,b)-(-1)^{m}\frac{a-a^{-1}}{2}e^{i\phi_n}Q(a,b)\\
    &\tilde{Q}(-a,b)=-(-1)^{m-1}\frac{a-a^{-1}}{2}e^{-i\phi_n}P(a,b)+(-1)^{m}\frac{a+a^{-1}}{2}e^{i\phi_n}Q(a,b)
\end{align}
Therefore, the statement is not proven, as it does not hold for the $n$th order (the $m$th order in $a$).

\section{Prospects of M-QSP}\label{sec3}
The inconsistencies discussed in sections \ref{sec1} and \ref{sec2} can be resolved by revising the conditions (ii) and (iii). Also, it is known that there exists a counterexample for Conjecture 2.1, as is pointed out in Ref. \cite{németh2023variants}:
\begin{align}
& P(a, b)=\frac{6}{25} \sqrt{\frac{37}{493}}\bigg[a^2 b^2+a^{-2} b^{-2}-\left(\frac{122}{37}+\frac{8 i}{37}\right)\left(b^2+b^{-2}\right)+\left(\frac{114}{37}+\frac{56 i}{37}\right)\left(a^{-2} b^2+a^2 b^{-2}\right)\nonumber\\
&~~~~~~~~~~~~~~~~~~~~~~~~~~~+\left(\frac{362}{111}-\frac{248 i}{111}\right)\left(a^2+a^{-2}\right)+\frac{692}{111}-\frac{719 i}{222}\bigg], \label{ce1}\\
& Q(a, b)=\frac{6}{25} \sqrt{\frac{37}{493}}\bigg[a^2 b^2-a^{-2} b^{-2}-\left(\frac{122}{37}+\frac{66 i}{37}\right)\left(b^2-b^{-2}\right)+\left(\frac{56}{37}+\frac{114 i}{37}\right)\left(a^{-2} b^2-a^2 b^{-2}\right)\nonumber\\
&~~~~~~~~~~~~~~~~~~~~~~~~~~~+\left(\frac{362}{111}-\frac{418 i}{111}\right)\left(a^2-a^{-2}\right)\bigg]\label{ce2}.
\end{align}
This means the conjecture is not a natural result of conditions (i)-(iv) and needs to be included as the condition (v). In this section, we prove these revised conditions make the necessary condition of M-QSP, where we provide the proof that follows the original paper as well as a simpler alternative. Moreover, we show that the complete reconstruction of the M-QSP Theorem is not straightforward. Because of the condition (v), ``$\Leftarrow$" of the theorem no longer holds, meaning some additional condition is required for the valid M-QSP Theorem.\\

\noindent{\bf Revised necessary condition of M-QSP:}
\textit{Let $n \in \mathbb{N}$. There exist $\Phi=\left\{\phi_0, \phi_1, \cdots, \phi_n\right\} \in \mathbb{R}^{n+1}$ and $s=\{s_1,\cdots,s_n\}\in\{0,1\}^n$ such that for all $(a, b) \in \mathbb{T}^2$ :
\begin{align}
    U_{(s, \Phi)}(a, b)=e^{i \phi_0 \sigma_z} \prod_{k=1}^n A^{s_k}\left(a\right) B^{1-s_k}\left(b\right) e^{i \phi_k \sigma_z}=\begin{pmatrix}
        P & Q \\
-Q^* & P^*
    \end{pmatrix},\label{Uphi}
\end{align}
where $A=\mathbb{I}\left(a+a^{-1}\right) / 2+\sigma_x\left(a-a^{-1}\right) / 2$ and $B=\mathbb{I}\left(b+b^{-1}\right) / 2+\sigma_x\left(b-b^{-1}\right) / 2$ \colorbox{yellow}{only if} $P, Q \in \mathbb{C}[a, b]$ (Laurent polynomials) and}
\begin{enumerate}
\item[(i)] \textit{$\operatorname{deg}(P) \preccurlyeq(m, n-m)$ and $\operatorname{deg}(Q) \preccurlyeq(m, n-m)$ for $m=|s|$ the Hamming weight of $s$.}
\item[(ii$^\prime$)] \textit{$P$ has \colorbox{yellow}{even parity} under $(a, b) \mapsto\left(a^{-1}, b^{-1}\right)$ and $Q$ has \colorbox{yellow}{odd parity} under $(a, b) \mapsto\left(a^{-1}, b^{-1}\right)$.}
\item[(iii$^\prime$)] \textit{$P$ has parity $m(\bmod 2)$ under $a \mapsto-a$ and parity $m-n(\bmod 2)$ under $b \mapsto-b$ and $\mathrm{Q}$ has parity \colorbox{yellow}{$m(\bmod 2)$} under $a \mapsto-a$ and parity \colorbox{yellow}{$n-m(\bmod 2)$} under $b \mapsto-b$.}
\item[(iv)] \textit{For all $(a, b) \in \mathbb{T}^2$ the relation $|P|^2+|Q|^2=1$ holds.}
\item[(v$^\prime$)] \textit{\colorbox{yellow}{For $m\ge1$ and $n-m\ge1$, $P_m(b)=e^{2i\varphi}Q_m(b)$ and/or $P_{n-m}(a)=e^{2i\varphi'}Q_{n-m}(a)$}, where $\varphi,\varphi'\in\mathbb{R}$ and $P_m(b)$ denotes the single variable Laurent polynomial coefficient of the highest $a^m$ term of $P$, and the same goes for others.}\\\par
\end{enumerate}
\noindent{\it Proof.} The following proof by induction focuses on changing the degree of $a$, but the same discussion can be applied to $b$. For $n=0$, $P=e^{i\phi_0}$ and $Q=0$ satisfy (i), (ii$^\prime$), (iii$^\prime$), and (iv), where (v$^\prime$) is not relevant for this degree. For $n=k-1,m=l-1$, assume the statement holds.
\begin{align}
    &U_{(\tilde{s},\tilde{\Phi})}=
    \begin{pmatrix}
        \tilde{P}&\tilde{Q}\\
        -\tilde{Q}^*&\tilde{P}^*
    \end{pmatrix}\\
    &\operatorname{deg}(\tilde{P}) \preccurlyeq(l-1, k-l)\\
    &\operatorname{deg}(\tilde{Q}) \preccurlyeq(l-1,k-l)\\
    &\tilde{P}(a^{-1},b^{-1})=\tilde{P}(a,b)\\
    &\tilde{Q}(a^{-1},b^{-1})=-\tilde{Q}(a,b)\\
    &\tilde{P}(-a,b)=(-1)^{l-1}\tilde{P}(a,b)\\
    &\tilde{Q}(-a,b)=(-1)^{l-1}\tilde{Q}(a,b)
\end{align}
For $n=k,m=l$, the sequence is calculated as
\begin{align}
    {U}_{(s,\Phi)}&\equiv
    \begin{pmatrix}
        P&Q\\
        -Q^*&P^*
    \end{pmatrix}\nonumber\\
    &=
    \begin{pmatrix}
        \tilde{P}&\tilde{Q}\\
        -\tilde{Q}^*&\tilde{P}^*
    \end{pmatrix}Ae^{i\phi_k\sigma_z}\nonumber\\    
    &=
    \begin{pmatrix}
    e^{i \phi_k}\left(\frac{a+a^{-1}}{2}\tilde{P}+\frac{a-a^{-1}}{2} \tilde{Q}\right)&e^{-i \phi_k}\left(\frac{a-a^{-1}}{2} \tilde{P}+\frac{a+a^{-1}}{2} \tilde{Q}\right)\\
    \cdot&\cdot
    \end{pmatrix},\label{Uphi_r}
\end{align}
where $\Phi=(\tilde{\Phi},\phi_k)$ and $s=(\tilde{s},1)$, and $P$ and $Q$ satisfy
\begin{align}
    &\operatorname{deg}(P) \preccurlyeq(l,k-l),\\
    &\operatorname{deg}(Q) \preccurlyeq(l,k-l),\\
    &P(a^{-1},b^{-1})=P(a,b),\\
    &Q(a^{-1},b^{-1})=-Q(a,b),\\
    &P(-a,b)=(-1)^{l}P(a,b),\\
    &Q(-a,b)=(-1)^{l}Q(a,b).
\end{align}
Also, $|P|^2+|Q|^2=1$ is satisfied because of the unitarity of the operations. If we take a look at the highest terms in $a$ in Eq. \eqref{Uphi_r}, we have
\begin{align}
    &P_l(b)=\frac{e^{i\phi_k}}{2}\left(\tilde{P}_{l-1}(b)+\tilde{Q}_{l-1}(b)\right),\\
    &Q_l(b)=\frac{e^{-i\phi_k}}{2}\left(\tilde{P}_{l-1}(b)+\tilde{Q}_{l-1}(b)\right),
\end{align}
which satisfy $P_l(b)=e^{2i\phi_k}Q_l(b)$. Therefore, the statement holds.\hfill $\square$\\\par

The same can be proven without using induction, so we show the alternate proof as well.\\

\noindent{\it Alternate Proof.} (i) is obviously satisfied, as the maximal degree of $a$ and $b$ are the number of applications of $A$ and $B$, respectively. Inversion parity of (ii$^\prime$) can be seen by noting the transformation of $a\mapsto a^{-1}$ and $b\mapsto b^{-1}$ are equivalent to $A\mapsto \sigma_zA\sigma_z$ and $B\mapsto \sigma_zB\sigma_z$, respectively. Under these transformations,
\begin{align}
    U_{(s,\Phi)}(a^{-1},b^{-1})&=e^{i \phi_0 \sigma_z} \prod_{k=1}^n \left(\sigma_zA^{s_k}\left(a\right)\sigma_z\right) \left(\sigma_zB^{1-s_k}\left(b\right)\sigma_z\right) e^{i \phi_k \sigma_z}\nonumber\\
    &=\sigma_zU_{(s,\Phi)}(a,b)\sigma_z\nonumber\\
    &=\begin{pmatrix}
        P(a,b)&-Q(a,b)\\
        Q^*(a,b)&P^*(a,b)
    \end{pmatrix}.
\end{align}
Comparing this with Eq. \eqref{Uphi}, we get
\begin{align}
    &P(a^{-1},b^{-1})=P(a,b),\\
    &Q(a^{-1},b^{-1})=-Q(a,b).
\end{align}
(iii$^\prime$) can also be checked easily by observing $a\mapsto -a$ and $b\mapsto -b$ correspond to $A\mapsto -A$ and $B\mapsto -B$, respectively. By simply counting the number of $A$ and $B$ in Eq. \eqref{Uphi}, we see
\begin{align}
    &P(-a,b)=(-1)^mP(a,b),\\
    &P(a,-b)=(-1)^{n-m}P(a,b),\\
    &Q(-a,b)=(-1)^mQ(a,b),\\
    &Q(a,-b)=(-1)^{n-m}Q(a,b).
\end{align}
(iv) is also satisfied because of the unitarity of the operations. To see (v$^\prime$), define
\begin{align}
    U_{(\tilde{s},\tilde{\Phi})}=e^{i \phi_0 \sigma_z} \prod_{k=1}^{n-1} A^{s_k}\left(a\right) B^{1-s_k}\left(b\right) e^{i \phi_k \sigma_z}=\begin{pmatrix}
        \tilde{P} & \tilde{Q} \\
        -\tilde{Q}^* & \tilde{P}^*
    \end{pmatrix},
\end{align}
then $U_{(s,\Phi)}$ can be computed as 
\begin{align}
    U_{(s,\Phi)}=U_{(\tilde{s},\tilde{\Phi})}A^{s_n}B^{1-s_n}e^{i\phi_n\sigma_z}.
\end{align}
When $s_n=1$, we have the following relation on the highest terms in $a$:
\begin{align}
    &P_m(b)=\frac{e^{i\phi_n}}{2}\left(\tilde{P}_{m-1}(b)+\tilde{Q}_{m-1}(b)\right),\\
    &Q_m(b)=\frac{e^{-i\phi_n}}{2}\left(\tilde{P}_{m-1}(b)+\tilde{Q}_{m-1}(b)\right).
\end{align}
These satisfy $P_m(b)=e^{2i\phi_n}Q_m(b)$. When $s_n=0$, we have $P_{n-m}(a)=e^{2i\phi_n}Q_{n-m}(a)$ from the similar discussion, so (v$^\prime$) is satisfied.
\hfill $\square$\\

However, these conditions (i)-(v$^\prime$) are actually not sufficient, i.e. ``$\Leftarrow$" of the M-QSP Theorem does not hold just with this revision. We show this by a counterexample, where conditions (i)-(v$^\prime$) are satisfied but no $U_{(s,\Phi)}$ can be constructed. Note that this counterexample is constructed based on the counterexample for Conjecture 2.1, but this proves the insufficiency of the conditions, which is another non-trivial task.

Define polynomials $\tilde{P}$ and $\tilde{Q}$ for $n=4$ and $m=2$ as in Eq. \eqref{ce1} and Eq. \eqref{ce2}, which satisfy conditions (i), (ii$^\prime$), (iii$^\prime$), and (iv) but not (v$^\prime$). Note that the coefficients of $\tilde{P}$ and $\tilde{Q}$ are all nonzero except for $\tilde{Q}_{00}$, the constant term of $\tilde{Q}$, which is restricted to zero by the condition (ii$^\prime$). Taking the contraposition of the revised necessary conditions of M-QSP, there exists no $U_{(\tilde{s},\tilde{\Phi})}$ for $\tilde{P}$ and $\tilde{Q}$. We construct $P$ and $Q$ for $n=5$ and $m=3$ by Eq. \eqref{Uphi_r} with a phase angle $\varphi$.
\begin{align}
    \begin{pmatrix}
        P&Q\\
        -Q^*&P^*
    \end{pmatrix}\equiv
    \begin{pmatrix}
        \tilde{P}&\tilde{Q}\\
        -\tilde{Q}^*&\tilde{P}^*
    \end{pmatrix}Ae^{i\varphi\sigma_z}
\end{align}
Because $\tilde{P}$ and $\tilde{Q}$ satisfy (i), (ii$^\prime$), (iii$^\prime$), and (iv), $P$ and $Q$ also satisfy (i), (ii$^\prime$), (iii$^\prime$), and (iv). Regarding (v$^\prime$), we have
\begin{align}
    &{P}_3(b)=\frac{e^{i\varphi}}{2}(\tilde{P}_2(b)+\tilde{Q}_2(b)),\\
    &{Q}_3(b)=\frac{e^{-i\varphi}}{2}(\tilde{P}_2(b)+\tilde{Q}_2(b)),
\end{align}
so $P_3(b)=e^{2i\varphi}Q_3(b)$ is satisfied. If we look at the highest and lowest degrees in $b$, this suggests $P_{32}=e^{2i\varphi}Q_{32}$ and $P_{3-2}=e^{2i\varphi}Q_{3-2}$. Using (ii$^\prime$), the latter relationship can be restated in terms of the highest degree in $b$, and we have $P_{32}=e^{2i\varphi}Q_{32}$ and $P_{-32}=-e^{2i\varphi}Q_{-32}$. Because these coefficients are all nonzero, $P_2(a)= e^{2i\varphi'}Q_2(a)$ cannot be satisfied. Therefore, $P$ and $Q$ satisfy (v$^\prime$) only for the highest order terms of $a$. Now we assume that there exists $U_{(s,\Phi)}$ for these $P$ and $Q$, which satisfy (i)-(v$^\prime$). 
\begin{align}
    \begin{pmatrix}
        {P} & {Q} \\
        -{Q}^* & {P}^*
    \end{pmatrix}=U_{(s,\Phi)}
\end{align}
Because of $P_3(b)=e^{2i\varphi}Q_3(b)$ and $P_2(a)\ne e^{2i\varphi'}Q_2(a)$, the last operator needs to be $A$ rather than $B$. Then we can express this $U_{(s,\Phi)}$ as
\begin{align}
    U_{(s,\Phi)}=\left(e^{i \phi_0 \sigma_z} \prod_{k=1}^4 A^{s_k}\left(a\right) B^{1-s_k}\left(b\right) e^{i \phi_k \sigma_z}\right)Ae^{i\phi_5\sigma_z}.
\end{align}
Rewriting the equation as
\begin{align}
    U_{(s,\Phi)}e^{-i\phi_5\sigma_z}A^\dag=e^{i \phi_0 \sigma_z} \prod_{k=1}^4 A^{s_k}\left(a\right) B^{1-s_k}\left(b\right) e^{i \phi_k \sigma_z},\label{tilde2}
\end{align}
we see the highest degree term of $a$ on the left-hand side needs to vanish so that both sides have the same degree. Calculating the left-hand side as
\begin{align}
    U_{(s,\Phi)}e^{-i\phi_5\sigma_z}A^\dag
    &=
    \begin{pmatrix}
        P&Q\\
        -Q^*&P^*
    \end{pmatrix}e^{-i\phi_5\sigma_z}A^\dag\nonumber\\
    &=
    \begin{pmatrix}
        \frac{a+a^{-1}}{2}e^{-i\phi_5}P-\frac{a-a^{-1}}{2}e^{i\phi_5}Q&-\frac{a-a^{-1}}{2}e^{-i\phi_5}P+\frac{a+a^{-1}}{2}e^{i\phi_5}Q\\
        \cdot&\cdot
    \end{pmatrix},
\end{align}
we get the condition on the highest degree terms of $a$ as
\begin{align}
    e^{-i\phi_5}P_3(b)-e^{i\phi_5}Q_3(b)=e^{-i\phi_5}P_3(b)\left(1-e^{2i(\phi_5-\varphi)}\right)=0.
\end{align}
Therefore, $\phi_5$ is either $\varphi$ or $\varphi+\pi$. When $\phi_5=\varphi$, Eq. \eqref{tilde2} becomes
\begin{align}
    U_{(s,\Phi)}e^{-i\varphi\sigma_z}A^\dag
    &=\begin{pmatrix}
        \tilde{P} & \tilde{Q} \\
        -\tilde{Q}^* & \tilde{P}^*
    \end{pmatrix}\nonumber\\
    &=e^{i \phi_0 \sigma_z} \prod_{k=1}^4 A^{s_k}\left(a\right) B^{1-s_k}\left(b\right) e^{i \phi_k \sigma_z},
\end{align}
meaning $\tilde{P}$ and $\tilde{Q}$ have $U_{(\tilde{s},\tilde{\Phi})}$. This contradicts with the fact that $\tilde{P}$ and $\tilde{Q}$ do not have $U_{(\tilde{s},\tilde{\Phi})}$. Similarly, when $\phi_5=\varphi+\pi$,
\begin{align}
    U_{(s,\Phi)}e^{-i(\varphi+\pi)\sigma_z}A^\dag
    &=\begin{pmatrix}
        -\tilde{P} & -\tilde{Q} \\
        \tilde{Q}^* & -\tilde{P}^*
    \end{pmatrix}\nonumber\\
    &=e^{i \phi_0 \sigma_z} \prod_{k=1}^4 A^{s_k}\left(a\right) B^{1-s_k}\left(b\right) e^{i \phi_k \sigma_z}
\end{align}
is also inconsistent with the assumption. Thus, there exists no $U_{(s,\Phi)}$ for these $P$ and $Q$ satisfying (i)-(v$^\prime$).

Therefore, the revised necessary conditions (i)-(v$^\prime$) are not sufficient for the existence of $U_{(s,\Phi)}$. To construct the valid M-QSP Theorem, some additional condition is required.

\section{Conclusion}\label{sec4}
We discussed inconsistencies appearing in the original paper of M-QSP \cite{Rossi2022multivariable} and provided a revised theorem on polynomials implementable with M-QSP. After modifying the inconsistencies, it turned out the currently known conditions are in fact only necessary conditions and not sufficient.
Under this circumstance, the important future research is to find necessary and sufficient conditions for M-QSP as ones for QSP \cite[Theorem 3]{Gily_n_2019}. Another key direction is to identify, even partly, the polynomials to which M-QSP is applicable. 

\subsection{Developments after preprint release}
After the appearance of this paper, many variants of M-QSP have been proposed, but unfortunately, the necessary and sufficient conditions are still not clarified in any protocol. In Ref. \cite{németh2023variants}, they use 
$W_z$ convention (e.g. signal operators are expressed as $Z$-rotation) and rotation operators are generalized to SU(2), where only necessary conditions are revealed. Similarly in Ref. \cite{laneve2024multivariatepolynomialsachievablequantum}, they employ an extended version of $W_z$ convention and they use SU(3) operator as rotation operators. Here, they give sufficient but not necessary conditions, and it is unclear how these conditions can be translated into the original M-QSP framework. Ref. \cite{gomes2024multivariableqspbosonicquantum} develops their protocol as the iterated version of standard QSP in $W_x$ convention (e.g. signal operators are expressed as $X$-rotation), where they also provide the application to bosonic quantum simulation. 

Compared with these methods, the original M-QSP framework has the advantage in its simplicity: it can be characterized by phase angles and ordering parameters, and the implementation is straightforward. 
As its potential application, it can provide an efficient algorithm for quantum state preparation, which is a task to prepare a state with a specific function encoded in the amplitudes \cite{mori2024state}.
Toward the actual application, Ref. \cite{ito2024polytime} proposes a decision program to determine if a given pair of polynomials $(P,Q)$ is implementable with M-QSP. However, any practical example of a polynomial $P$ that is implementable with M-QSP nor the way to construct the complementary polynomial $Q$ is not known yet.
\section*{Acknowledgement}
This work is supported by MEXT Quantum Leap Flagship Program (MEXTQLEAP) Grant No. JPMXS0120319794, and JST COI-NEXT Program Grant No. JPMJPF2014.

\bibliographystyle{quantum}
\bibliography{cite}

\end{document}